# Design Space and Variability Analysis of SOI MOSFET for Ultra-Low Power Band-to-Band Tunneling Neurons

Jay Sonawane[1], Shubham Patil[1], Abhishek Kadam[1], Ajay Kumar Singh[1], Sandip Lashkare[2], Veeresh Deshpande[1], Udayan Ganguly[1]

*Abstract*— Large spiking neural networks (SNNs) require ultra-low power and low variability hardware for neuromorphic computing applications. Recently, a band-to-band tunneling-based (BTBT) integrator, enabling sub-kHz operation of neurons with area and energy efficiency, was proposed. For an ultra-low power implementation of such neurons, a very low BTBT current is needed, so minimizing current without degrading neuronal properties is essential. Low variability is needed in the ultra-low current integrator to avoid network performance degradation in a large BTBT neuron-based SNN. To address this, we conducted design space and variability analysis in TCAD, utilizing a well-calibrated TCAD deck with experimental data from GlobalFoundries 32nm PD-SOI MOSFET. First, we discuss the physics-based explanation of the tunneling mechanism. Second, we explore the impact of device design parameters on SOI MOSFET performance, highlighting parameter sensitivities to tunneling current. With device parameters' optimization, we demonstrate a ~20× reduction in BTBT current compared to the experimental data. Finally, a variability analysis that includes the effects of random dopant fluctuations (RDF), oxide thickness variability (OTV), and channel-oxide interface traps ($D_{IT}$) in the BTBT, SS, and ON regimes of operation is shown. The BTBT regime shows high sensitivity to the RDF and OTV as any variation in them directly modulates the tunnel length or the electric field at the drain-channel junction, whereas minimal sensitivity to $D_{IT}$ is observed.

## I. INTRODUCTION

Neuromorphic computing architectures consisting of biologically inspired neural networks are excellently manifested by the spiking neural networks (SNNs) [1]. The enormous number of neurons in the human brain makes it essential to use large SNNs to realize biological neural networks. Compact area and energy-efficient technology are essential for large-scale deployment of neuromorphic computing hardware for applications such as speech processing, the Internet of Things, and sensors [2]. The advanced technological nodes are optimized for area and power efficiency from a CMOS logic application perspective [3]. However, the neuron implementation in the ON-regime (ON) suffers from higher power dissipation and large area

This work was supported in part by the Department of Science and Technology (DST), India, the Indian Institute of Technology Bombay Nano-Fabrication (IITBNF) facility IIT Bombay, and the Centre for Excellence (CEN), IIT Bombay.
Jay Sonawane, Shubham Patil, Veeresh Deshpande, and Udayan Ganguly are with the Department of Electrical Engineering, Indian Institute of Technology Bombay, India (e-mail: jay.sonawane.iitb@gmail.com, udayan@ee.iitb.ac.in).
Sandip Lashkare is with the Department of Electrical Engineering, IIT Gandhinagar, Gujarat 382355, India (e-mail: lashkaresandip@gmail.com).
The authors would like to acknowledge Prof. Nihar Ranjan Mohapatra for the discussion.

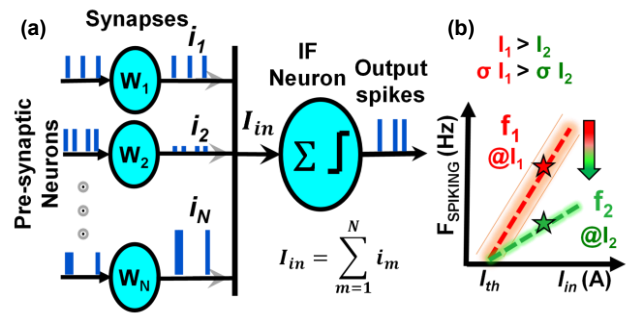

Fig. 1. **SNN implementation.** (a) Schematic of SNN with spiking neurons and plastic synapses. (b) High variability in synaptic current translates to high variability in neuron performance. Hence, low current and low variability are essential for large-scale SNN implementations without substantial degradation in network performance.

requirements. Further, the ON-based CMOS neurons require a dedicated on-chip capacitor [4]–[9]. In attempts to reduce the power dissipation and capacitor size, the subthreshold (SS) regime-based neuron was proposed [9], [10]. However, the SS regime is highly susceptible to the PVT variability [11]. In a quest to find the area and energy-efficient technology to realize SNNs, a band-to-band tunneling (BTBT) regime-based neuron was proposed using the GlobalFoundries (GF) 45nm Partially Depleted Silicon-on-Insulator (PD-SOI) technology. The body capacitor made it a compact design, and the low-current operation facilitated energy efficiency [1], [12]. This technology was based on standard SOI technology, so further integration was supported naturally. The BTBT-based compact area and energy-efficient low-power neurons displayed a successful hardware implementation [12]. While BTBT-based neuron operation has been successfully implemented in SNN hardware, minimizing the BTBT current becomes essential for achieving ultra-low power operation of the neuron, especially in large SNN implementations. Therefore, it is necessary to comprehensively explore and analyze the device design space and optimize the device to reduce BTBT current. The lower current can decelerate the BTBT-induced hole-storage, thereby decreasing spiking frequency for an enhanced biologically plausible neuron implementation. Neurons like the Olfactory Bulb, Basal Ganglia, and Brainstem [13] have sub-kHz spiking frequency, and hence, for hardware realization of neurons like these, low current, energy-efficient neuron operation is essential.

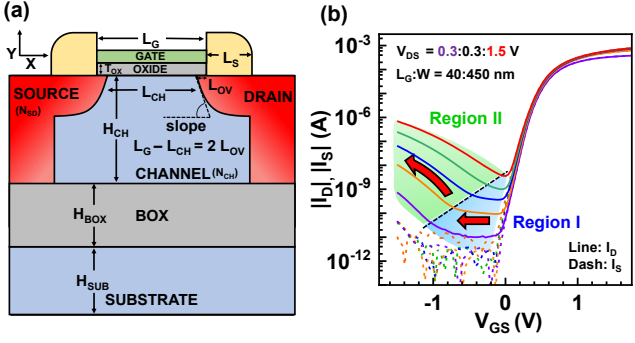

Fig. 2. **SOI MOSFET structure and experimental transfer characteristics.** (a) The schematic representation of the PD-SOI MOSFET used in this work. (b) Experimentally measured drain ($I_D$) and source ($I_S$) current as a function of gate voltage ($V_{GS}$) for different drain biases ($V_{DS}$) for the GlobalFoundries 32nm PD-SOI MOSFET [17]–[19]. Two different current slopes are observed in the quantum tunneling regime indicated as blue (Region I) and green regions (Region II).

Further, an optimized device structure should not affect the performance in ON and SS regimes to facilitate hardware implementation in the same technology. Reduction in BTBT current in high mobility ultra-thin strained cap-Si/$Si_xGe_{1-x}$ heterostructure on SOI form logic application perspective had been reported earlier using the cap-Si thickness and $Si_xGe_{1-x}$ composition and thickness engineering [14-16]. However, a comprehensive analysis of Si channel MOSFETs pertaining to the BTBT regime operation, underlying physics, and device design dependence has not been explored.

Furthermore, device-to-device variability is a crucial factor while designing a neuron in addition to low current. In these systems, synaptic currents integrate to charge the leaky capacitor, generating a membrane potential. Once this potential exceeds a predefined threshold, the neuron spikes (Fig. 1). Variations in integration current result in variability in the integration time constant, leading to inconsistent spiking frequencies. This neuron-to-neuron variability significantly impacts network performance. Therefore, low variability is indeed needed in the ultra-low current integrator to prevent network performance degradation. Therefore, it is essential to analyze the potential impact of process variability in the BTBT regime from a technological standpoint.

In this work, we have characterized the 32nm PD-SOI MOSFET and analyzed the BTBT physics using a well-calibrated Sentaurus TCAD deck [17]. Next, we performed an exploratory analysis in the design space on the calibrated deck. We showed a ~20× reduction in BTBT current by optimizing the device parameters without sufficiently degrading the SS and ON regime performance. Finally, the impact of process variability, i.e., random dopant fluctuations (RDF), oxide thickness variability (OTV), and channel-oxide interface traps ($D_{IT}$) in the BTBT, SS, and ON regimes of operation, is illustrated.

## II. DEVICE DETAILS

Fig. 2(a) shows the schematic cross-section of a PD-SOI MOSFET used in this work with the device dimension abbreviations. The 32nm PD-SOI MOSFET fabricated at GF with gate length ($L_G$)/width (W) of 40/450 nm is used in this work [18]–[20]. The electrical measurements are performed using an Agilent B1500 Semiconductor parameter analyzer. A well-calibrated Sentaurus TCAD deck is used to understand the physics behind the tunneling mechanism, design space, and variability analysis. For design space analysis, variation in channel height ($H_{CH}$), buried oxide thickness ($H_{BOX}$), substrate height ($H_{SUB}$), drain-gate (D-G) and source-gate (S-G) overlap length ($L_{OV}$), equivalent oxide thickness ($T_{OX}$), channel ($N_{CH}$) and source/drain doping ($N_{SD}$) is considered. Variation in the RDF, OTV, and $D_{IT}$ is considered for variability analysis. The benchmarking of the BTBT regime is performed with the SS and ON regimes. The current in BTBT, SS, and ON regimes are extracted at V$_{GS}$ of -1.5, 0.25, and 1.2 V, respectively, at fixed V$_{DS}$ of 0.9 V.

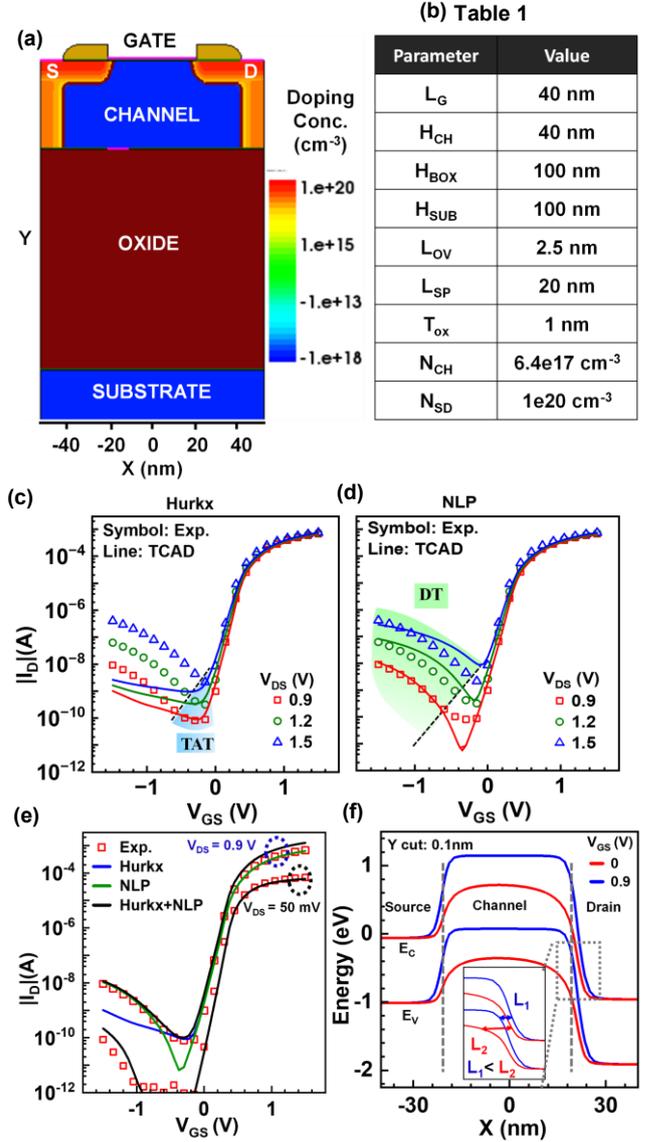

Fig. 3. **BTBT physics.** (a) Device parameters used in the calibrated structure. (b) Table 1 summarizes the device parameters used in TCAD calibration. (c) Simulation using the Hurkx model matches with the experimental data at low $|V_{GS}|$ (dominated by TAT). (d) The Nonlocal path model matches with the experimental data at high $|V_{GS}|$ (dominated by DT). However, it fails to capture TAT adequately at low $|V_{GS}|$. (e) The addition of NLP and the Hurkx model perfectly matches experimental data with TCAD simulations in the BTBT regime. (f) The energy band diagram (EBD) profile (with Y-cut at 0.1 nm below gate oxide) for two different values of $V_{GS}$ (0 V and 0.9 V) shows two distinct tunnel lengths. A low tunnel length manifested by a high electric field facilitates DT, and TAT is dominant in the case of a higher tunnel length.

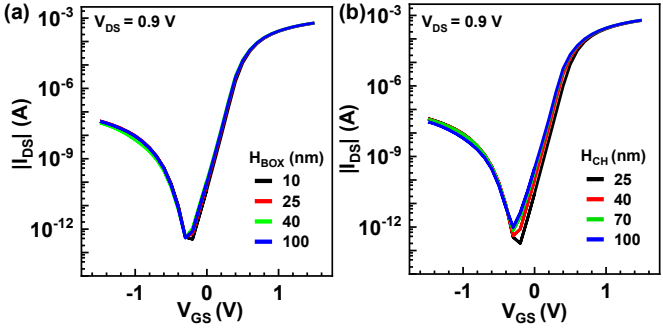

Fig. 4. **H$_{BOX}$ and H$_{CH}$ variation impact**. (a-b) Simulated transfer characteristics (I$_{DS}$ vs. V$_{GS}$) on varying box height (H$_{BOX}$) and channel thickness (H$_{CH}$). H$_{BOX}$ and H$_{CH}$ variation shows negligible impact in the BTBT regime.

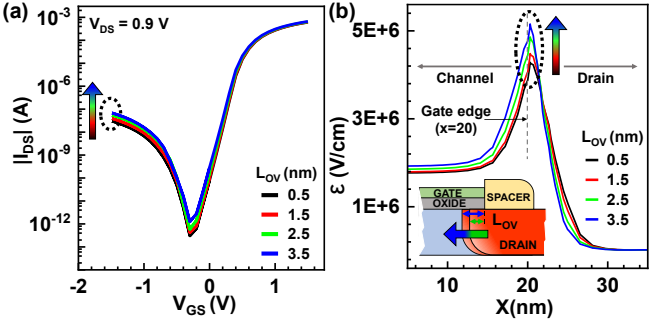

Fig. 5. **L$_{OV}$ variation impact**. (a) Simulated transfer characteristics (I$_{DS}$ vs. V$_{GS}$) with L$_{OV}$ variation. The BTBT current increases on increasing L$_{OV}$. (b) Electric field in the D-G overlap region (x = 10 nm to 30 nm, Y-cut 0.1 nm below gate oxide) obtained from TCAD simulation. An increase in the electric field within the D-G overlap region is observed as L$_{OV}$ increases, resulting in a higher BTBT current.

### III. RESULTS AND DISCUSSION

#### A. Experimental characterization

Fig. 2(b) shows the experimentally measured drain (I$_D$) and source current (I$_S$) as a function of gate voltage (V$_{GS}$) at different drain biases (V$_{DS}$) for the GlobalFoundries 32nm PD-SOI MOSFET. The presence of band-to-band tunneling at the drain-channel junction is evident from the increase in I$_D$ with drain and gate bias. Two different current slopes are observed in the tunneling regime, indicated as blue regions (Region I) and green regions (Region II). We will identify and analyze the bias dependence of different tunneling mechanisms in the following subsection.

#### B. Device Calibration and Physics

Fig. 3(a) shows the schematic cross-section of the PD-SOI used in the Sentaurus TCAD. To calibrate the Sentaurus TCAD deck, the device dimensions used are (a) gate length (L$_G$) of 40 nm, (b) channel height (H$_{CH}$) of 40 nm, (c) buried oxide thickness (H$_{BOX}$) of 100 nm, (d) substrate height (H$_{SUB}$) of 100 nm, (e) D-G and S-G overlap length (L$_{OV}$) of 2.5 nm, (f) spacer length (L$_{SP}$) of 20 nm, (g) Equivalent oxide thickness (T$_{OX}$) of 1 nm, (h) channel doping (N$_{CH}$) of 6.4e17 cm$^{-3}$, and (i) source/drain doping (N$_{SD}$) of 1e20 cm$^{-3}$ (Table

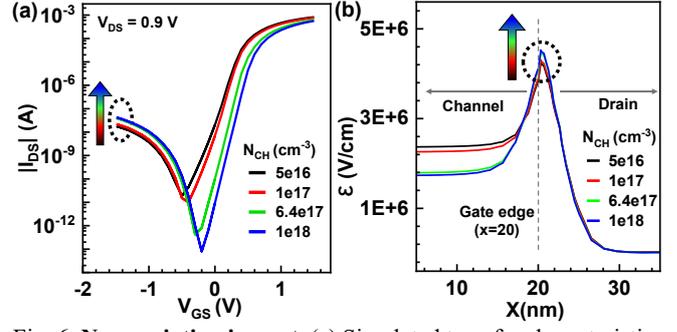

Fig. 6. **N$_{CH}$ variation impact**. (a) Simulated transfer characteristics (I$_{DS}$ vs. V$_{GS}$) and (b) electric field in the D-G overlap region (x = 10 nm to 30 nm, Y-cut 0.1 nm below gate oxide) with channel doping (N$_{CH}$) variation. An increase in the electric field within the D-G overlap region is observed with increases in N$_{CH}$, resulting in higher BTBT current.

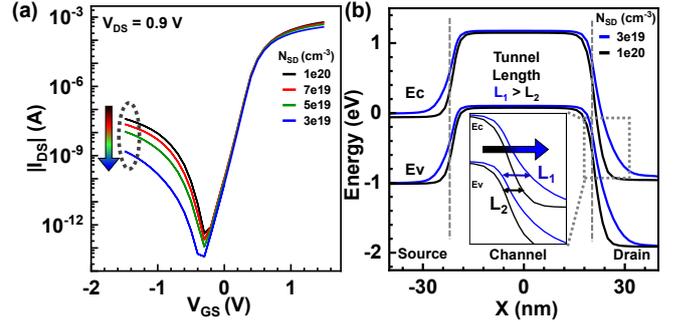

Fig. 7. **N$_{SD}$ variation impact.** (a) Simulated transfer characteristics (I$_{DS}$ vs V$_{GS}$) on varying S-D doping (N$_{SD}$). On decreasing N$_{SD}$, BTBT current decreases. (b) The across the channel (x = -40 nm to 40 nm, Y-cut 0.1 nm below gate oxide) obtained from TCAD simulation. Higher N$_{SD}$ results in lower tunnel-length, resulting in higher BTBT current.

I). To capture the device physics, the following models are used: (1) the drift-diffusion model to represent current transport, (2) doping-dependent mobility models (Masetti and PhuMob) and the Lombardi model to capture mobility degradation, (3) the extended Canali model to account for high field saturation, and (4) SRH and Auger models to represent recombination-generation processes. Dynamic non-local path (NLP) tunneling and the Hurkx model [21] are employed to capture the tunneling phenomenon. By utilizing these models, an excellent match is achieved in the SS and ON regimes (Fig. 3(e)). It is imperative to analyze the nature of tunneling for proper calibration, as discussed in the quantum tunneling regime (Fig. 2(b)). In literature, the dominance of TAT (at low bias) before DT sets in at high bias has been reported in tunnel field-effect transistors (TFETs) [22]-[24]. Similarly, to understand the tunneling behavior in the GF 32nm SOI MOSFET, we used Hurkx and NLP separately to characterize the presence of TAT and DT dominant regions, respectively. Fig. 3(c) shows that the Hurkx model matches the experimentally obtained current at lower |V$_{GS}$|. However, a significant mismatch at higher |V$_{GS}$| is observed. In contrast, the NLP model captures the BTBT at a higher |V$_{GS}$|. However, it shows a mismatch at lower |V$_{GS}$| (Fig. 3(d)). After adding the contributions from both models, the Sentaurus TCAD showed an excellent match with the experimental data in the BTBT regime (Fig. 3(e)). The energy-band diagram (EBD) is plotted along a horizontal cut-line taken 0.1 nm below the Si-SiO$_2$ interface for two different gate biases (V$_{GS}$) of 0 V and 0.9 V at a fixed V$_{DS}$ of

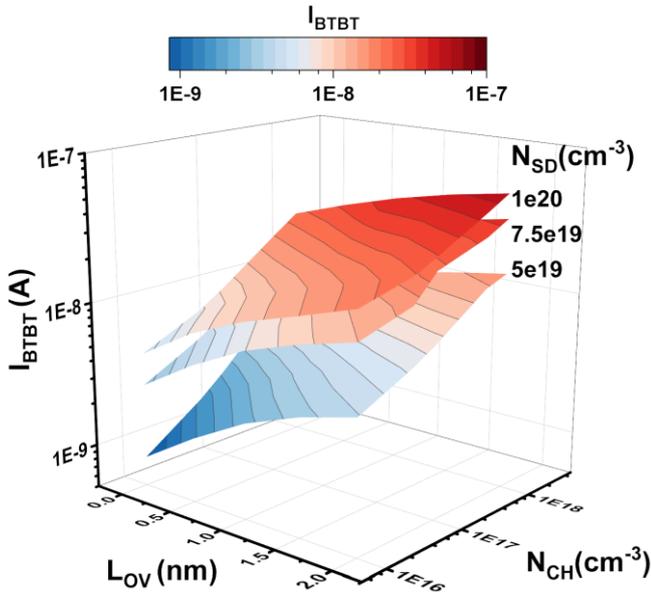

Fig. 8. **BTBT regime design space.** The BTBT design space 3D surface colormap showing $I_{BTBT}$ for different $N_{SD}$, $N_{CH}$, and $L_{OV}$ values is plotted ($V_{GS}$ = -1.5 V and $V_{DS}$ = 0.9 V). For a value of $N_{SD}$ (1e20, 7.5e19, and 5e19), $I_{BTBT}$ obtained on varying the $L_{OV}$ and $N_{CH}$ is plotted. We observe that $I_{BTBT}$ increases with an increase in $N_{SD}$, $N_{CH}$, and $L_{OV}$.

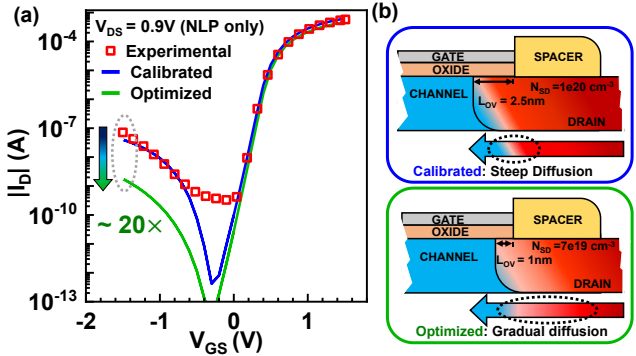

Fig. 9. **BTBT regime optimization and optimized device.** (a) Simulated transfer characteristics ($I_{DS}$ vs. $V_{GS}$) of the optimized device. Optimizing the device for BTBT operation leads to a ~20× decrease in the BTBT current at $V_{GS}$ = -1.5 V, with a 20% decrease in ON current. (b) The optimized device structure uses a gradual S/D extension diffusion and lower $N_{SD}$ doping, resulting in a longer tunnel length and lower BTBT current.

0.9 V (Fig. 3(f)). We observed longer tunnel lengths at a lower gate bias, and at a higher gate bias, we observed shorter tunnel lengths. The nature of the tunnel length (long, short) is manifested by the electric field (low, high) due to the bias (low, high). Hence, the current in the quantum tunneling regime that is dominated by TAT at lower |$V_{GS}$| is modeled using the Hurkx model (Region I, Fig. 2(b)), and the current dominated by DT at higher |$V_{GS}$| is modeled using the NLP model (Region II, Fig. 2(b)). The approximate effective voltage ($V_{DS}$ + |$V_{GS}$|) added up at the point where DT starts is consistent throughout the sweeps for different $V_{DS}$. As we are interested in the DT current, the NLP model (capturing DT) is used for further study in this work. This well-calibrated deck is then used for the detailed device design space and variability analysis.

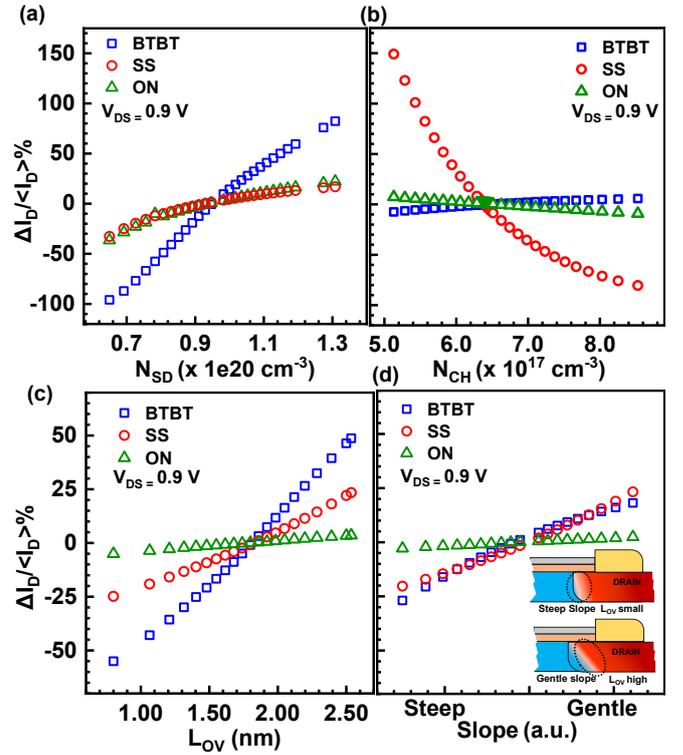

Fig. 10. **RDF variability.** Sensitivity of the BTBT, SS and ON regime currents expressed as percentage change w.r.t. (a) $N_{SD}$ (b) $N_{CH}$ (c) $L_{OV}$, and (d) S/D extension diffusion slope variation (inset shows schematic of gentle and steep slope of S/D extension in the D-G overlap region). For analysis, the BTBT, SS and ON regime currents are extracted at $V_{GS}$ of -1.5 V, 0.25 V, and 1.2 V at fixed $V_{DS}$ of 0.9 V. The BTBT regime shows maximum sensitivity to $N_{SD}$ compared to the SS and ON regime whereas low sensitivity to the $N_{CH}$ and $L_{OV}$ variation. BTBT and SS regime shows similar sensitivity to the S/D extension doping profile slope variability.

C. Device design space analysis and optimization for the BTBT Regime

   i. Device design space analysis

To understand the impact of the device's physical and process parameters on the BTBT current, a design space analysis was done. The effect of variation in the $H_{CH}$, $H_{BOX}$, $L_{OV}$, $N_{CH}$, and $N_{SD}$ was studied using the transfer characteristics at a fixed $V_{DS}$ of 0.9 V.

Figs. 4(a) and 4(b) illustrate the impact of variation in $H_{BOX}$ and $H_{CH}$ scaling on the transfer characteristics, respectively. Notably, our investigation reveals that these variations result in negligible changes in the BTBT current. This is primarily because BTBT predominantly occurs in the D-G overlap region. Variations in $H_{BOX}$ and $H_{CH}$ have little impact on the electric field dynamics and energy band profile in this region. Consequently, the BTBT carrier generation remains essentially unchanged, resulting in consistent BTBT current levels.

In Fig. 5(a), a slight increase in the BTBT current is observed as the D-G overlap ($L_{OV}$) increases. This observation is attributed to the enhanced effective coupling between the drain and gate biases (due to the increased overlap area), resulting in an elevated electric field at the junction, as depicted in Fig. 5(b). We observe that the effective electric field increases with an increase in $L_{OV}$. This higher electric field ultimately translates into an increase in the BTBT current.

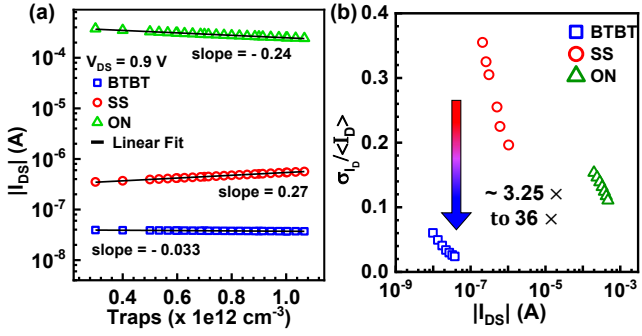

Fig. 11. **Channel-oxide interface traps variability.** (a) Extracted BTBT, SS and ON regime current from simulated transfer characteristics at $V_{GS}$ of -1.5, 0.25, and 1.2 V, respectively, at fixed $V_{DS}$ of 0.9 V as a function of channel-oxide interface trap concentration. SS regime shows maximum sensitivity to the increased trap concentration, whereas BTBT regime shows negligible dependence. (b) Calculated coefficient of variation ($\sigma_{I_D}/<I_D>$) as a function of the drain current ($I_D$). The BTBT regime shows ~3.25× to 36× lower variability than SS regime.

Fig. 6(a) shows that BTBT current increases with an increase in the channel doping ($N_{CH}$). A higher $N_{CH}$ value leads to a reduction in the depletion width at the drain-channel junction. This reduction in depletion width results in a steeper energy band profile, shorter tunnel lengths, and a higher electric field in the overlap region (Fig. 6 (b)). We observe that the electric field increases a little as $N_{CH}$ increases. These factors collectively enhance the probability of direct tunneling events, consequently contributing to the increased BTBT current.

In Fig. 7(a), a notable increase in the tunneling current is observed with an increase in the S-D doping ($N_{SD}$). Fig. 7(b) illustrates the EBD for two different doping profiles along the channel, with a horizontal cut-line taken 1 nm below the Si-SiO₂ interface. Higher $N_{SD}$ reduces the depletion width at the junction, resulting in a shorter tunnel length and an increased electric field, ultimately leading to a higher BTBT current.

The design space analysis highlights that the dependence of BTBT current can be attributed to $L_{OV}$, $N_{CH}$, and $N_{SD}$. Fig. 8 shows a 3D surface colormap of $I_{BTBT}$ while varying $L_{OV}$ and $N_{CH}$ for 3 values of $N_{SD}$ (5e19, 7.5e19, and 1e20 cm⁻³). This figure summarizes the design parameters required for a specific $I_{BTBT}$ and illustrates the dependence of each parameter on $I_{BTBT}$.

### ii. Optimization for BTBT Regime Operation

The optimization strategy for reducing direct tunneling without significantly affecting the OFF current, SS, and ON regime current through the design space exploration study can be summarized as decreasing $N_{CH}$, $N_{SD}$, and $L_{OV}$. However, the decrease in $N_{CH}$ significantly affects the OFF current and the subthreshold slope (SS); hence, it can't be perturbated significantly. Hence, by using a lower $L_{OV}$ (1 nm) and lower $N_{SD}$ (7e19 cm⁻³) with a gradual extension doping profile, we achieve a ~20 × reduction in the BTBT current, as depicted in Fig. 9 (a). Fig. 9 (b) illustrates the schematic cross-section comparison between the calibrated and optimized device structure used in TCAD simulations for qualitative understanding. The lower BTBT current can translate to lower spiking frequency, leading to an energy-efficient realization of SNN hardware.

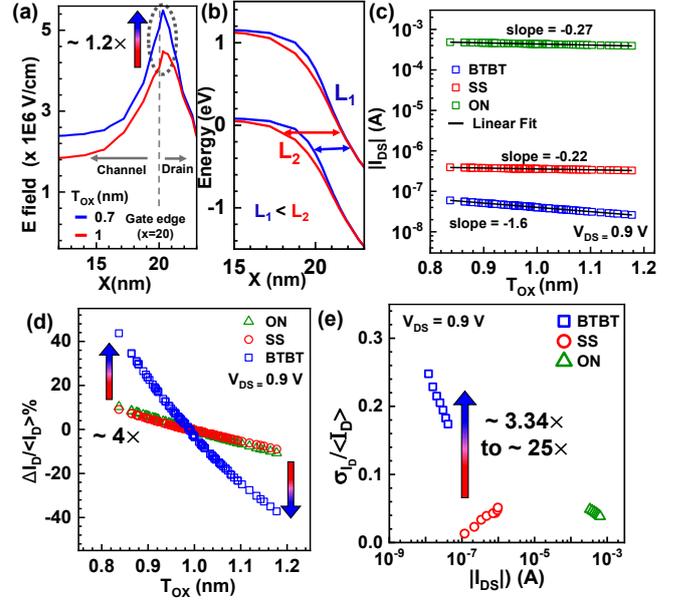

Fig. 12. **OTV variability.** (a) Electric field across the channel and (b) EBD (Y-cut 0.1 nm below gate oxide) for two different oxide thickness. Variability in $T_{OX}$ affects the tunnel length and the electric field at the junction. (c) Extracted BTBT, SS and ON regime current from simulated transfer characteristics at $V_{GS}$ of -1.5, 0.25, and 1.2 V, respectively, at fixed $V_{DS}$ of 0.9 V, as a function of oxide thickness variation. The BTBT regime exhibits the highest sensitivity, followed by the ON and SS regimes. (d) Sensitivity of $I_D$ vs. $T_{OX}$ for the three regimes shows that BTBT regime exhibit higher variability compared to the SS regime. (e) The CV as a function of the $I_D$ comparison in the three operating regimes shows that the BTBT regime exhibits ~3.34× to ~25× higher CV than the SS regime for same current range resulting in higher variability in BTBT regime current.

### D. VARIABILITY ANALYSIS

In addition to a very low current, low variability is a significant concern in the design of low-power circuits. This is essential because the performance impact and complexities of compensating for variability add extra costs. Therefore, it is essential to analyze the process parameters that influence the BTBT regime variability and determine the extent of their impact. This comparison of the BTBT regime is conducted alongside the SS and ON regimes. The impact of variation in the RDF, OTV, and $D_{IT}$ is discussed in the next subsection. For variability analysis, the percentage change ($\frac{\Delta I_D}{<I_D>}$) % and coefficient of variation (CV) are compared among these three regimes.

$$\text{Percentage change} = \frac{\Delta I_D}{<I_D>}\%$$

$$CV = \frac{\sigma_{I_D}}{<I_D>}$$

Where, $\sigma_{I_D}$ is the standard deviation, $<I_D>$ is the mean drain current, and $\Delta I_D = I_D - <I_D>$.

### i. Random Dopant Fluctuations (RDF)

The variations in four parameters, i.e., $N_{CH}$, $N_{SD}$, $L_{OV}$, and S/D extension slope, are considered to understand the RDF impact. For analysis, the BTBT, SS, and ON regime currents are extracted at $V_{GS}$ of -1.5 V, 0.25 V, and 1.2 V at fixed $V_{DS}$ of 0.9 V, using the simulated transfer characteristics. A $V_{GS}$ of 0.25 V is used for subthreshold regime variability simulation as the current lies sufficiently above the threshold current for spiking to occur [25]. Fig. 10(a-d) shows the sensitivities measured by percentage change comparison among the three regimes for $N_{SD}$, $N_{CH}$, $L_{OV}$, and S/D extension slope variation, respectively. The BTBT regime shows maximum sensitivity to $N_{SD}$, compared to the SS and ON regime (Fig. 10(a)), whereas low sensitivity to the $N_{CH}$ (Fig. 10(b)) and $L_{OV}$ (Fig. 10(c)) variation. The BTBT current's lower sensitivity on $N_{CH}$ than $N_{SD}$ is because $N_{CH}$ is approximately 2 to 3 orders lower than $N_{SD}$ concentration. Hence, the nature of the energy band diagram is dominated by $N_{SD}$. The slope of the S/D extension is defined as the steepness of the S/D extension doping profile with respect to a horizontal axis parallel to the gate oxide, as shown in Fig. 2(a). SS and BTBT regime currents show similar sensitivity when the slope is varied (Fig. 10(d)). A schematic demonstrating a steep and gentle slope is shown in the inset of Fig. 10(d). The SS regime is highly sensitive to slope changes as the effective channel length decreases, introducing short channel effects. The sensitivity of the BTBT regime to the slope can be attributed to an increase in the D-G overlap area, resulting in a greater area with a higher electric field, which is suitable for higher tunnel carrier generation.

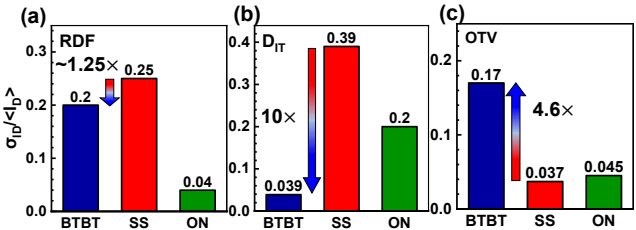

Fig. 13. **Variability benchmarking.** Comparison of CV for the BTBT, SS, and ON Regimes at $V_{GS}$ values of -1.5, 0.25 and 1.2 V, respectively, at fixed $V_{DS}$ of 0.9 V for (a) RDF, (b) $D_{IT}$, and (c) OTV variability. BTBT regime shows ~1.25×, and ~10× lower variability compared to SS regimes for RDF and $D_{IT}$ variability, respectively. However, ~4.6× higher variability is observe in BTBT regimes compared to the SS for OTV.

   ii.   Channel-Oxide Interface Traps ($D_{IT}$)

To investigate the sensitivity towards traps, donor-type traps are added at the channel-gate oxide interface. The trap states are included at the midgap in simulations. Fig. 11(a) shows the variation in the BTBT, SS, and ON regime current extracted at $V_{GS}$ of -1.5 V, 0.25 V, and 1.2 V, respectively, at fixed $V_{DS}$ of 0.9 V using simulated transfer characteristics for varying trap concentrations. Notably, the BTBT regime demonstrates minimal sensitivity (slope) to the introduced traps, whereas the SS followed by the ON regime exhibits the highest sensitivity. The presence of interface traps significantly impacts several important parameters, including threshold voltage ($V_T$), subthreshold slope (SS), and channel mobility. Consequently, this influence leads to substantial variability in the SS and ON regimes. However, it's essential to note that interface traps do not significantly affect the tunnel length and electric field at the drain-channel junction, where BTBT occurs. Therefore, the BTBT regime demonstrates low sensitivity to these traps [26]. The calculated CV shows that the BTBT regime offers ~3.25× to ~36× lower variability as compared to the SS regime for a similar range of current (Fig. 11(b)).

   iii.   Oxide thickness variation (OTV)

Although foundry processes are the most optimized to counter the variations in gate oxide thickness, the OTV study is essential to demonstrate oxide thickness sensitivity towards the BTBT regime. Fig. 12(a) shows the electric field profile for $T_{OX} = 0.7$ and 1 nm along the channel-drain interface, 0.1nm below the gate oxide. A higher electric field in and lower tunnel length (Fig. 12 (b)) is observed in the case of a low $T_{OX}$. Fig. 12 (c) shows the variation in the BTBT, SS, and ON regime currents extracted at $V_{GS}$ of -1.5 V, 0.25 V, and 1.2 V at fixed $V_{DS}$ of 0.9 V for varying $T_{OX}$. It is observed that the BTBT regime exhibits the highest sensitivity to OTV variability, followed by the ON and SS regimes. The calculated change in $I_D$ vs. $T_{OX}$ for the three regimes shows that the BTBT regime exhibits higher variability than the SS regime (Fig. 12(d)). The calculated CV as a function of the $I_D$ comparison in the three operating regimes shows that the BTBT regime exhibits ~3.34× to ~25× higher CV than the SS regime for the same current range (Fig. 12(e)). The high sensitivity of BTBT can be attributed to the fact that any change in the oxide thickness results in changes in the effective coupling between the drain and gate, especially within the overlap region. These alterations influence the energy band profile within the device, subsequently affecting both the electric field within the channel-drain junction and the tunnel length. Both of these factors play crucial roles in regulating the BTBT current. Although this study involves a variability study with substantial variation in $T_{OX}$, we expect to observe minimal effect of $T_{OX}$ variability in the state-of-the-art optimized foundry processes.

E. Variability Summary

To summarize, CV comparison of the BTBT, SS, and ON regime at $V_{GS}$ of -1.5, 0.25, and 1.2 V, respectively, at fixed $V_{DS}$ of 0.9 V is shown for (a) RDF, (b) $D_{IT}$, and (C) OTV variability (Fig. 13). BTBT regime shows ~1.25× and ~10× lower variability for RDF and $D_{IT}$ compared the SS regimes. However, the BTBT regime showed high sensitivity to the OTV variation. This highlights that RDF and $D_{IT}$ have a relatively minor impact on the BTBT regime, while OTV variation has a more pronounced influence.

   IV.   CONCLUSION

We performed design space and variability analysis to achieve a lower BTBT current. To achieve higher energy efficiency and lower spiking frequency in BTBT-based neurons, we first investigated the underlying physics in the BTBT regime. We confirmed the dominance of TAT at lower gate bias and DT at higher gate bias using the well-calibrated TCAD deck. Second, we infer from the simulations that a low $L_{OV}$, low $N_{SD}$, and low $N_{CH}$ are favorable parameters for reducing the BTBT current. We demonstrated an optimized device structure using a smaller $L_{OV}$, lower $N_{SD}$, and gradual S/D extension doping, leading to a ~20× reduction in the BTBT current at the same operating bias. Further, we performed a variability study for BTBT current due to RDF, $D_{IT}$, and OTV. We showed that BTBT current is most sensitive to OTV and has comparable variability as that of subthreshold current due to RDF. Hence, our work presents the essential design parameters to achieve an ultra-low current BTBT-based neuron to enable real-time neuron implementation akin to biological neurons, along with a variability study to demonstrate relative sensitivities towards BTBT, SS, and ON regime.